\documentclass[conference]{IEEEtran}
\usepackage{cite}
\usepackage{amsmath,amssymb,amsfonts}
\usepackage{algorithmic}
\usepackage{graphicx}
\usepackage{algorithm,algorithmic}
\usepackage{hyperref}
\hypersetup{hidelinks=True}
\usepackage{textcomp}
\usepackage{multirow}
\usepackage{multicol}
\usepackage{booktabs}
\usepackage{tikz}
\usepackage{adjustbox}
\usepackage{amsmath}
\usepackage{amsfonts}
\usepackage{mathtools}
\usepackage{tikz}
\usepackage{pifont}
\usetikzlibrary{calc}
\newcommand{\cmark}{\ding{51}}%
\newcommand{\xmark}{\ding{55}}%
\newcommand\new[1]{\textcolor{black}{#1}}

\begin{document}
\title{MLVICX: Multi-Level Variance-Covariance Exploration for Chest X-ray Self-Supervised Representation Learning}

\author{Azad Singh, \IEEEmembership{Student Member, IEEE}, Vandan Gorade and Deepak Mishra, \IEEEmembership{Member, IEEE}
\thanks{ Azad Singh and Deepak Mishra are with the Indian Institute of Technology, Jodhpur, 342030 India (e-mail: singh.63@iitj.ac.in, dmishra@iitj.ac.in). }
\thanks{Vandan Gorade is with Savitribai Phule Pune University, Pune, 411007 India. At the time of this work, he was an intern 
at the Department of Computer Science and Engineering, Indian Institute of Technology, Jodhpur, 342030, India (e-mail: vangorade@gmail.com).}}
\maketitle

\begin{abstract}
Self-supervised learning (SSL) is potentially useful in reducing the need for manual annotation and making deep learning models accessible for medical image analysis tasks. By leveraging the representations learned from unlabeled data, self-supervised models perform well on tasks that require little to no fine-tuning. \new{However, for medical images, like chest X-rays, which are characterized by complex anatomical structures and diverse clinical conditions, there arises a  need for representation learning techniques that can encode fine-grained details while preserving the broader contextual information.}
In this context, we introduce MLVICX (Multi-Level Variance-Covariance Exploration for Chest X-ray Self-Supervised Representation Learning), an approach to \new{capture rich representations in the form of embeddings from chest X-ray images.} Central to our approach is a novel multi-level variance and covariance exploration strategy that empowers the model to \new{effectively detect} diagnostically meaningful patterns while \new{reducing} redundancy. By enhancing the variance and covariance of the learned embeddings, MLVICX promotes the retention of critical \new{medical insights} by adapting both global and \new{local contextual} details. We demonstrate the \new{performance} of MLVICX in advancing self-supervised chest X-ray representation learning through comprehensive experiments. The performance enhancements we observe across various downstream tasks \new{highlight} the significance of the proposed approach in enhancing the utility of chest X-ray embeddings for precision medical diagnosis and comprehensive image analysis. For pertaining, we used the NIH-Chest X-ray dataset, while for downstream tasks, we utilized NIH-Chest X-ray, Vinbig-CXR, RSNA pneumonia, and SIIM-ACR Pneumothorax datasets. Overall, we observe more than 3\% performance gains over \new{SOTA SSL approaches} in various downstream tasks.  
\end{abstract}


\section{Introduction}
\label{sec:introduction}
\IEEEPARstart{M}{edical} image analysis is a critical component of modern healthcare, enabling clinicians to diagnose diseases accurately, plan treatment strategies, and monitor patient progress. Traditionally, \new{it requires} laborious manual identification and analysis of features within images, leading to subjectivity and time constraints. The advent of deep learning and AI has \new{opened up} a new era of possibilities, addressing these challenges through automated feature extraction from medical images, ultimately aiding the diagnosis process~\cite{zhang2022cxr,wang2019thorax}. While deep learning models have demonstrated remarkable potential in various medical image analysis tasks, their performance is \new{governed by} the quality and diversity of the labeled data used for training. Acquiring accurate labels for medical images is a labor-intensive task that demands specialized expertise, accurate annotation, and significant time investment~\cite{krishnan2022self,chen2022recent}.

Self-supervised learning (SSL)~\cite{simclr, moco,byol,simsiam} offers \new{an alternative} approach to mitigate the challenges associated with acquiring labeled data. It has gained popularity in the field of medical image analysis due to its ability to learn representations with limited annotated data~\cite{huang2022self,lu2023two,10032792}. SSL capitalizes on the idea that images inherently contain abundant latent information, which can serve as a surrogate for traditional explicit labels. This allows SSL algorithms to design pretext tasks, where the model learns to predict certain aspects of the data without requiring external annotations.
As the model \new{attempts} to solve these pretext tasks, it learns to extract relevant features directly from the images. Many studies have shown the effectiveness of SSL in various medical image analysis tasks, including classification~\cite{sowrirajan2021mococxr,vu2021medaug,azizi2021big}, detection~\cite{xie2022unimiss}, and segmentation~\cite{chaitanya2020contrastive,kalapos2023self,ouyang2020self}. These recent advancements predominantly leverage discrimination learning, where models extract features from input images to differentiate between variations while remaining resilient to transformations. Training models to distinguish between different variations of the same input, such as different viewpoints, lighting conditions, or augmentations, enables capturing \new{of} meaningful, invariant representations.

In the modern SSL context, discrimination techniques fall into two main categories: contrastive and non-contrastive methods, which \new{widely adopt} the joint-embedding framework, often realized through Siamese network architecture\new{~\cite{jing2020self}}. Contrastive SSL methods bring embeddings of positive pairs closer together while pushing them away from embeddings of negative pairs, where \new{one way of creating} the positive pairs \new{involves using} augmented versions of the same image while negative pairs \new{consists of} augmented versions of different images\new{~\cite{tian2020understanding}}. On the other hand, non-contrastive methods also bring together embeddings of positive pairs but avoid the use of explicit negative pairs, either by choosing some specific architectural design choices or by adopting explicit regularization techniques\new{~\cite{tian2021understanding}}.

Nonetheless, conventional SSL~\cite{simclr,simclrv2,byol,vicreg} techniques have been adapted and applied to medical image analysis tasks.
However, certain limitations emerge that hinder the quality of learned representations while effectively navigating the complexities inherent to medical images, particularly chest X-rays.
One notable challenge lies in the effective fusion of coarse and fine-grained details within \new{chest X-rays}. \new{Chest X-rays encompass a wide spectrum of information, ranging from subtle anomalies like pulmonary nodules or infiltrates to anatomical features such as the heart silhouette or lung contours. For instance, in the case of pulmonary nodules, an accurate diagnosis requires the localization of small, subtle lesions within the lung tissue. These anomalies are often challenging to detect due to their size and location within the chest cavity. On the other hand, anatomical features like the heart silhouette or diaphragm shape are essential for understanding the overall clinical context of the patient's condition. Analyzing the heart silhouette can provide valuable information about the heart's size, shape, and position within the chest, which is essential for diagnosing heart conditions and assessing overall cardiac health. Meanwhile, examining lung contours can help detect abnormalities, such as masses or fluid accumulation within the lungs, and can also provide information about the lung's overall condition and position in the chest. Achieving a balance between capturing these subtle anomalies and preserving the broader anatomical context is critical to effective chest X-ray analysis.~\cite{woznitza2018chest,majkowska2020chest}}

\new{Most of the} existing SSL methods, while successful in capturing either global \new{features that provide an overview of the entire image} or \new{excel in capturing } fine-grained features \new{that might signify anomalies or abnormalities}, often struggle to holistically integrate both~\cite{ponti2017everything,simonyan2014very}.
This \new{disparity in feature extraction} leads to representations that either lack the broader global context or overlook the fine-grained details ~\cite{denseCL,woo2018cbam,pixelpro}. 

To address these limitations, we propose MLVICX (Multi-Level Variance-Covariance Exploration for Chest X-ray Self-Supervised Representation Learning). \new{MLVICX, through explicit regularization of variance and covariance in a multi-level framework, aims to fuse global and local contextual details within chest X-ray images.} 
To achieve this, MLVICX \new{uses the} intermediate feature maps \new{of the learned representations} from diverse layers of the backbone encoders. MLVICX empowers the model to capture feature representations of varying complexity and abstraction by progressively harnessing intermediate feature maps. Further, these feature maps are meticulously channeled through context-bottleneck blocks~\cite{woo2018cbam} to extract contextual information while \new{concurrently enhancing the features captured by the core encoder at multiple levels, resulting in locally guided global representation.} 
The hierarchical contextual information extracted through the context-bottleneck blocks is subsequently aggregated, encouraging fine-grained representations. 

\new{While previous approaches have investigated the extraction of both global and local features~\cite{pcrlv2,pcrlv1,intermoco}, they often do not explicitly tackle representation variability. These methods primarily focus on preserving a wide range of information in learned representations, while MLVICX introduces a multi-level variance-covariance exploration to regulate the utilization of global and local contextual details. Additionally, in PCRLv1 and PCRLv2, the incorporation of a reconstruction module introduces added computational complexity into the training process.} \new{By explicitly regulating features from multiple intermediate levels using variance and covariance terms, MLVICX enhances variability at multiple levels and adapts the utilization of global and local contextual details. This results not only in more contextually rich and informative representations but also inherently discriminative.}

We assess the quality of the representations acquired through our novel approach by conducting a comprehensive comparative analysis against SOTA self-supervised methods, including SimCLR~\cite{simclr}, MoCo~\cite{moco}, BYOL~\cite{byol}, VICReg~\cite{vicreg}, and Barlow-Twins~\cite{barlow}. Additionally, to provide an evaluation within the domain of chest X-ray analysis, we benchmark our approach against recently proposed methods specifically tailored for this context, namely Inter\_MoCo~\cite{intermoco}, PCRLv1~\cite{pcrlv1}, and PCRLv2~\cite{pcrlv2}. The results of our comprehensive evaluation consistently demonstrate a noteworthy enhancement in performance across all considered datasets, highlighting the effectiveness and stability of the representations generated by the proposed approach.

\section{Related Works}

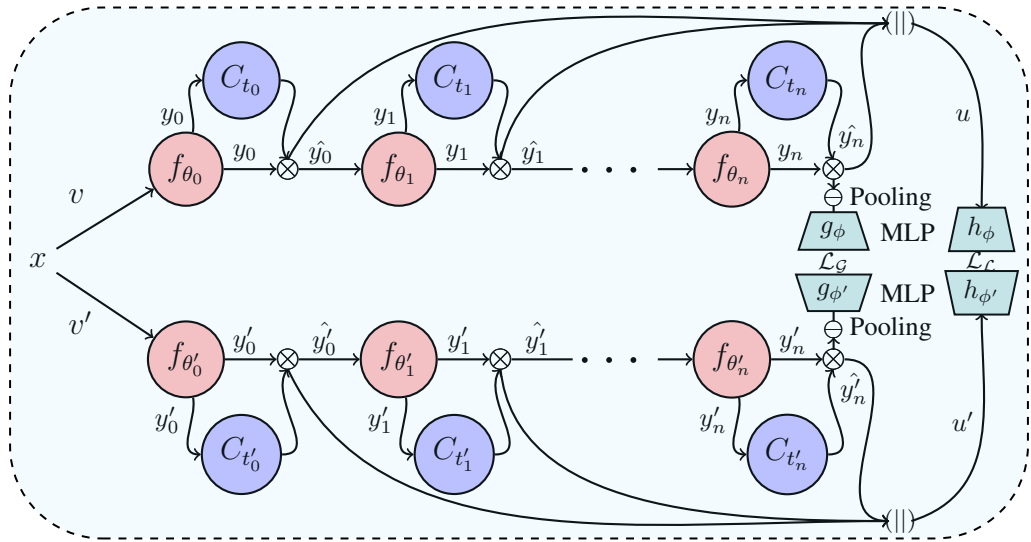
\begin{figure*}[htbp]
\centering
\tikzset{every picture/.style={line width=0.75pt}}
\usetikzlibrary{shapes}
\begin{adjustbox}{trim=0 0 0 0.6cm}%
\begin{tikzpicture}[x=1pt,y=1pt,yscale=0.6,xscale=0.7]

\draw (15,218.5) node [font=\large] (ip){$x$};

\draw (30,265) node [font=\large] [anchor=north west][inner sep=0.75pt]  {$v$};
\draw (30,170) node [font=\large] [anchor=south west] [inner sep=0.75pt] {$v'$};

\node[circle,thick,draw=black,text=black,fill=red!25,font=\large] (la1) at (95,276) {$f_{\theta_0}$};
\draw [->,thick,black] (ip) -- (la1);
\draw (127,286.5) node  (f1){$y_{0}$};
\draw (168,286.5) node  (f1){$\hat{y_0}$};

\node[circle,thick,draw=black,text=black,fill=blue!25,font=\large] (ua1) at (125,332) {$C_{t_0}$};
\draw [->,thick,black] (la1) to [out=80,in=180] (ua1);
\draw (88,309) node  (f1){$y_{0}$};
\coordinate (b1) at (144,276);
\draw (150,276) node [thick][font=\large] {${\otimes }$};
\draw (150,276) node [thick][font=\large] {${\otimes }$};
\draw (150,276) node [thick][font=\large] {${\otimes }$};
\draw [->,thick,black] (la1) -- (b1);

\coordinate (b2) at (150,282);
\draw [->,thick,black] (ua1) to [out=00,in=120] (b2);

\coordinate (b3) at (156,276);
\node[circle,thick,draw=black,text=black,fill=red!25,font=\large] (la2) at (210,276) {$f_{\theta_1}$};
\draw [->,thick,black] (b3) -- (la2);
\draw (242,286.5) node  (f1){$y_{1}$};
\draw (283,286.5) node  (f1){$\hat{y_1}$};

\node[circle,thick,draw=black,text=black,fill=blue!25,font=\large] (ua2) at (240,332) {$C_{t_1}$};
\draw [->,thick,black] (la2) to [out=80,in=180] (ua2);
\draw (203,309) node (f1){$y_{1}$};

\coordinate (b4) at (259,276);
\draw (265,276) node [thick][font=\large] {${\otimes }$};
\draw (265,276) node [thick][font=\large] {${\otimes }$};
\draw (265,276) node [thick][font=\large] {${\otimes }$};
\draw [->,thick,black] (la2) -- (b4);

\coordinate (b5) at (265,282);
\draw [->,thick,black] (ua2) to [out=00,in=120] (b5);
\coordinate (b6) at (271,276);
\coordinate (b7) at (304,276);
\draw [-,thick,black] (b6) -- (b7);
\draw (305,284) node [anchor=north west][inner sep=0.60pt][thick,black,font=\huge]  {$\cdots $};
\coordinate (b8) at (350,276);

\node[circle,thick,draw=black,text=black,fill=red!25,font=\large] (la3) at (390,276) {$f_{\theta_n}$};
\draw [->,thick,black] (b8) -- (la3);
\draw (422,286.5) node  (f1){$y_{n}$};
\draw (455,296.5) node  (f1){$\hat{y_n}$};

\node[circle,thick,draw=black,text=black,fill=blue!25,font=\large] (ua3) at (420,332) {$C_{t_n}$};
\draw [->,thick,black] (la3) to [out=80,in=180] (ua3);
\draw (383,309) node (f1){$y_{n}$};

\coordinate (b9) at (438,276);  
\draw (445,276) node [thick][font=\large] {${\otimes }$};
\draw (445,276) node [thick][font=\large] {${\otimes }$};
\draw (445,276) node [thick][font=\large] {${\otimes }$};
\draw [->,thick,black] (la3) -- (b9);
\coordinate (b10) at (445,282);
\draw [->,thick,black] (ua3) to [out=00,in=120] (b10);

\draw (445,258) node [ultra thick,text=black] {${\ominus }$};
\draw (445,258) node [ultra thick,text=black] {${\ominus }$};
\draw (445,258) node [ultra thick,text=black] {${\ominus }$};

\coordinate (b11) at (445,270);
\coordinate (b12) at (445,263);
\draw [->,thick,black] (b11) -- (b12);
\draw (485,236) node {MLP};
\draw (476,257) node {Pooling};
\coordinate (b13) at (445,253);
\coordinate (b14) at (445,248);
\draw [-,thick,black] (b13) -- (b14);

\node[trapezium,
    draw = black,
    text = black,
    fill = teal!20,
    minimum width = 5mm,
    minimum height = 5mm,
    trapezium angle = 72,] (t) at (445,237) {$g_\phi$};


\node[circle,thick,draw=black,text=black,fill=red!25,font=\large] (ta1) at (95,156) {$f_{\theta'_0}$};
\draw [->,thick,black] (ip) -- (ta1);
\draw (127,168) node  (f1){$y'_{0}$};
\draw (170,169) node  (f1){$\hat{y'_0}$};

\coordinate (c1) at (144,156);
\draw (150,156) node [thick][font=\large] {${\otimes }$};
\draw (150,156) node [thick][font=\large] {${\otimes }$};
\draw (150,156) node [thick][font=\large] {${\otimes }$};
\draw [->,thick,black] (ta1) -- (c1);

\node[circle,thick,draw=black,text=black,fill=blue!25,font=\large] (ka1) at (125,96) {$C_{t'_0}$};
\draw [->,thick,black] (ta1) to [out=-80,in=-180] (ka1);
\draw (85,118) node  (f1){$y'_{0}$};

\coordinate (c2) at (150,150);
\draw [->,thick,black] (ka1) to [out=00,in=-120] (c2);

\coordinate (c3) at (156,156);
\node[circle,thick,draw=black,text=black,fill=red!25,font=\large] (ta2) at (210,156) {$f_{\theta'_1}$};
\draw [->,thick,black] (c3) -- (ta2);
\draw (243,168) node  (f1){$y'_{1}$};
\draw (285,169) node  (f1){$\hat{y'_1}$};

\coordinate (c4) at (259,156);
\node[circle,thick,draw=black,text=black,fill=blue!25,font=\large] (ka2) at (240,96) {$C_{t'_1}$};
\draw [->,thick,black] (ta2) to [out=-80,in=-180] (ka2);
\draw (200,118) node  (f1){$y'_{1}$};

\draw (265,156) node [thick][font=\large] {${\otimes }$};
\draw (265,156) node [thick][font=\large] {${\otimes }$};
\draw (265,156) node [thick][font=\large] {${\otimes }$};

\draw [->,thick,black] (ta2) -- (c4);
\coordinate (c5) at (265,150);
\draw [->,thick,black] (ka2) to [out=00,in=-120] (c5);

\coordinate (c6) at (271,156);
\coordinate (c7) at (304,156);
\draw [-,thick,black] (c6) -- (c7);
\draw (305,164) node [anchor=north west][inner sep=0.60pt][thick,black,font=\huge]  {$\cdots $};

\coordinate (c8) at (350,156);
\node[circle,thick,draw=black,text=black,fill=red!25,font=\large] (ta3) at (390,156) {$f_{\theta'_n}$};
\draw [->,thick,black] (c8) -- (ta3);
\draw (423,168) node  (f1){$y'_{n}$};
\draw (456,138) node (f1){$\hat{y'_n}$};

\node[circle,thick,draw=black,text=black,fill=blue!25,font=\large] (ka3) at (420,96) {$C_{t'_n}$};
\draw [->,thick,black] (ta3) to [out=-80,in=-180] (ka3);
\draw (380,118) node (f1){$y'_{n}$};

\coordinate (c9) at (439,156);
\draw (445,156) node [thick][font=\large] {${\otimes }$};
\draw (445,156) node [thick][font=\large] {${\otimes }$};
\draw (445,156) node [thick][font=\large] {${\otimes }$};

\draw [->,thick,black] (ta3) -- (c9);
\coordinate (c10) at (445,150);
\draw [->,thick,black] (ka3) to [out=00,in=-120] (c10);

\draw (445,175) node [thick] {${\ominus }$};
\draw (445,175) node [thick] {${\ominus }$};
\draw (445,175) node [thick] {${\ominus }$};
\coordinate (c11) at (445,162);
\coordinate (c12) at (445,170);
\draw [->,thick,black] (c11) -- (c12);
\draw (476,175) node {Pooling};
\draw (485,198) node {MLP};
\draw (445,217) node [font=\small] {$\mathcal{L_G}$};
\draw (525,218) node [font=\small] {$\mathcal{L_L}$};

\coordinate (c13) at (445,180);
\coordinate (c14) at (445,187);
\draw [-,thick,black] (c13) -- (c14);
\node[trapezium,
    draw = black,
    text = black,
    fill = teal!20,
    minimum width = 1mm,
    minimum height = 1mm,
    trapezium angle = -72,] (t) at (445,198) {$g_{\phi'}$};

\draw (481,368) node [font=\small] {$( || )$};

\coordinate (o1) at (475,368);
\coordinate (o2) at (487,368);
\draw [->,thick,black] (b2) to [out=65,in=180] (o1);
\draw [->,thick,black] (b5) to [out=80,in=180] (o1);
\coordinate (b100) at (451,276);
\draw [->,thick,black] (b100) to [out=00,in=180] (o1);

\coordinate (f1) at (525,250);
\draw [->,thick,black] (o2) to [out=340,in=90] (f1);
\node[trapezium,
    draw = black,
    text = black,
    fill = teal!20,
    minimum width = 5mm,
    minimum height = 5mm,
    trapezium angle = 78,] (t) at (525,238) {$h_\phi$};

\draw (481,54) node [font=\small] {$( || )$};

\coordinate (o3) at (475,54);
\coordinate (o4) at (487,54);
\draw [->,thick,black] (c2) to [out=-68,in=-180] (o3);
\draw [->,thick,black] (c5) to [out=-85,in=-180] (o3);
\coordinate (c100) at (451,156);
\draw [->,thick,black] (c100) to [out=00,in=-180] (o3);

\draw (515,310) node {$u$};
\draw (515,117) node {$u'$};

\coordinate (f2) at (525,185);
\draw [->,thick,black] (o4) to [out=-340,in=-90] (f2);
\node[trapezium,
    draw = black,
    text = black,
    fill = teal!20,
    minimum width = 1mm,
    minimum height = 1mm,
    trapezium angle = -78,] (t) at (525,198) {$h_{\phi'}$};

\draw [dashed][rounded corners=35pt][color={rgb, 255:red, 00; green, 00; blue, 0 }  ,draw opacity=1][fill=cyan!40!white  ,fill opacity=0.1 ] (0,43) -- (550,43) -- (550,378) -- (0,378) -- cycle ;

\end{tikzpicture}
\end{adjustbox}
\caption{Architecture of the proposed approach. $v$ and $v'$ are two augmented versions of the input $x$, process by weight shared encoders $f_{\theta}$ and $f_{\theta{'}}$ respectively to give $y$ and $y'$. $C_{t_p}$ and $C_{t'_p}$ are the context-bottleneck blocks to refine the intermediate representations $y_p$ and $y'_p$. The refined intermediate representations are aggregated to make compound representations $u$ and $u'$  in each branch. $( || )$ represents the concatenation of multi-level intermediate feature maps, while $\otimes$ represents multiplication.}
\label{fig:overview}
\end{figure*}


\subsection{Self Supervised Learning}
SSL involves a two-step \new{training} process: pretraining and downstream training. \new{In conventional SSL} during pretraining, a model is trained on a diverse dataset using auxiliary tasks that are designed to generate supervisory signals from the data itself. These tasks encourage the model to capture useful patterns, structures, and relationships in the data, effectively learning informative features. Once the model is pretrained, the learned representations are fine-tuned \new{for the} downstream tasks in which a smaller labeled dataset is used to adapt the pretrained features to a specific task, such as classification, object detection, or segmentation. SSL offers \new{various}  pretraining options, where one approach involves designing hand-crafted pretext tasks~\cite{exemplecnn,zhuang2019self,zhu2020rubik}, such as \new{predicting} image rotations~\cite{rotpred} or contextually consistent patches, solving a jigsaw puzzle~\cite{jigsaw}, image colorization~\cite{colorization} \new{etc}. Another \new{possible approach} is instance discrimination methods,  wherein contrastive methods like \new{SimCLR~\cite{simclr}, MoCo~\cite{moco} etc.}, non-contrastive methods such as \new{BYOL~\cite{byol}, SimSiam~\cite{simsiam}, Barlow-Twins (BT)~\cite{barlow},VICReg~\cite{vicreg} etc.} are employed. These methods are designed to foster the acquisition of invariant image representations, thereby mitigating the impact of image transformations. Contrastive methods aim to bring embeddings of similar instances \new{(positive pairs)} closer while pushing apart embeddings of dissimilar instances \new{(negative pairs)}. Non-contrastive methods also emphasize similarity but avoid explicit negative pairs and can be further categorized into \new{self-}distillation-based approaches like BYOL~\cite{byol} and SimSiam~\cite{simsiam} and information maximization methods such as Barlow-Twins~\cite{barlow} and recently proposed VICReg~\cite{vicreg}. Another line of work in SSL involves clustering and generative methods where clustering-based~\cite{SwaV,caron2018deep,zhan2020online} tasks encourage models to group similar instances together, while generative tasks~\cite{parmar2021dual,chen2020generative} involve generating parts of an image and learning to reconstruct it. These diverse pretraining \new{approaches} harness the intrinsic information present in data, enabling models to acquire rich representations that can be fine-tuned for various downstream tasks, contributing to the effectiveness of SSL across a range of \new{vision applications}.

\subsection{SSL for \new{Chest X-ray Images}}
Given the inherent scarcity of precisely annotated medical data, SSL emerges as a compelling approach in this domain. Among its various methodologies, instance discrimination techniques, particularly contrastive approaches, have gained considerable \new{attention} and applicability in \new{chest X-ray images.}
For instance, Hari \textit{et al.} introduced MoCo-CXR~\cite{sowrirajan2021mococxr}, an 
adaptation of the MoCo~\cite{moco} framework tailored for harnessing meaningful representations from chest X-ray images. Drawing inspiration from SimCLR~\cite{simclr}, Azizi \textit{et al.} presented an approach \new{called} as Multi-Instance Contrastive Learning (MICLe) in~\cite{azizi2021big}. This method leverages the power of multiple patient images to construct positive pairs, resulting in \new{better representations compared to SimCLR}. \new{Similarly} MedAug~\cite{vu2021medaug} introduced a novel strategy for selecting positive pairs across varying views \new{using metadata}. A \new{interesting work called as} DiRA~\cite{dira}, introduced by Haghighi \textit{et al.}, which unites discriminative, restorative, and adversarial learning to collaboratively \new{learn} complementary visual information from unlabeled medical images for fine-grained semantic representation learning. In~\cite{pcrlv1} Zhou \textit{et al.} proposed preservational contrastive representation learning (PCRL) for both X-ray and CT modalities that enrich learned medical representations by dynamically reconstructing diverse image contexts. In~\cite{pcrlv2}, authors introduced PCRLv2, which is an updated version of PCRL. PCRLv2 is a unified framework addressing pixel-level restoration and scale information, achieving \new{better} results across diverse tasks and datasets. In~\cite{intermoco}, Kaku \textit{et al.} enhances contrastive learning with intermediate-layer closeness beyond just the final layer, demonstrating improved feature quality over standard MoCo. \new{While these approaches showed benefits, however, they commonly overlooked the explicit handling of representation variability. In the context of chest X-ray analysis, representation variability is important due to the commonality of anatomical structures across patients. Without it, the model might generalize too broadly, while missing the subtle abnormalities. In contrast, MLVICX is uniquely tailored to address representational variability effectively.} It leverages a multi-level framework to progressively enhance variance and covariance, resulting in refined feature maps from diverse layers. This approach captures fine-grained local details and overarching semantic concepts.

\section{Methodology}

Fig.~\ref{fig:overview} provides the overview of the proposed SSL framework, which is based on the siamese configuration of the deep neural network. The following subsections provide details on the particulars of MLVICX. 
\subsection{\new{Proposed} SSL Framework}
As shown in Fig.~\ref{fig:overview}, the proposed SSL framework employs a joint embedding architecture that incorporates two branches, each consisting of a backbone CNN encoder ($f_\theta$ and $f_{\theta'}$ \new{respectively}) followed by global MLP heads ($g_\phi$ and $g_{\phi'}$) and \new{multi-level} MLP heads ($h_\phi$ and $h_{\phi'}$). The input image, $x \in \mathcal{D}$ where $\mathcal{D}$ is the training set, undergoes random image transformations $t$ and $t'$ to generate two perturbed variants $v$ and $v'$. These transformations encompass both semantic-level operations such as flipping, rotation, and random cropping, as well as pixel-level adjustments like gaussian blur, color jittering, and solarization. The views $v$ and $v'$ are \new{passes through} $f_\theta$ and $f_{\theta'}$ to generate global latent representations, denoted as $y = f_\theta(v)$ and $y' = f_{\theta'}(v')$ respectively.

\subsubsection{Intermediate Representations and Context-Bottleneck Heads} As the core of the MLVICX, \new{is to} extracts intermediate representations to encode contextual information from varying layers of complexity. These intermediate representations play a pivotal role in \new{integration of} global and local \new{contextual} details within chest X-rays. The encoders ($f_\theta$ and $f_{\theta'}$) generate intermediate feature maps at different levels, referred to as $y_p \new{\in {\{y_0,y_1,........,y_n\}}}$ and $y'_p \new{\in {\{y'_0,y'_1,........,y'_n\}}}$ respectively, where $p$ corresponds to the \new{$p_{th}$} layer of the encoder. The dimensions of $y_p$ and $y'_p$ are denoted as $c_p \times h_p \times w_p$, reflecting the number of channels $c_p$ along with the height $h_p$ and width $w_p$. These intermediate feature maps $y_p$ and $y'_p$ are \new{used} by context-bottleneck heads ($C_{t_p}$ and $C_{t'_p}$) \new{respectively.}

The context-bottleneck heads consist of the channel pooling block and the spatial pooling block. The channel pooling block takes $y_p$ as input and produces a 1D channel-activated map $a^c_p$ $\in \mathbb{R}^{c_p \times 1 \times 1}$ according to the formulation:
\begin{equation}
a^{ch}_{p} = \sigma(Conv_1(AvgPool(y_p)) + Conv_1(MaxPool(y_p)))
\label{cam}
\end{equation}
Here, $\sigma$ represents the sigmoid activation function. The channel-activated map $a^{ch}_p$ is then \new{modulated} with $y_p$ using element-wise multiplication to yield the feature map $k^{ch}_p = (a^{ch}_{p} \otimes y_p)$ $\in \mathbb{R}^{c_p \times h_p \times w_p}$. \new{Additionally, $Conv_1$ refers to a convolutional layer with a $1 \times 1$ kernel size.}
The spatial pooling block is complementary to the channel pooling block. It takes the feature map $k^c_{p}$ as input and generates a 2D spatially-activated map $a^s_p$ $\in \mathbb{R}^{1 \times h_p \times w_p}$ using the formulation:
\begin{equation}
a^s_{p} = \sigma(Conv_7([AvgPool(k^{ch}_{p});MaxPool(k^{ch}_{p})]))
\label{sam}
\end{equation} 
Similarly, the spatially-activated map $a^s_p$ is \new{modulated} with $k^{ch}_{p}$ using element-wise multiplication to yield the refined feature map $\hat{y_p} = (a^s_{p} \otimes k^{ch}_{p})$. The process described above is symmetrically applied to obtain $\hat{y'_p}$, the refined feature map from the other branch. 
The refined intermediate representations $\hat{y_p}$ and ${\hat{y'_p}}$ are processed by subsequent layers of the encoder, which enables the model to learn locally guided context-aware global representations $y$ and $y'$. 
\new{Subsequently for} the multi-level variance-covariance exploration, we combine $\hat{y_p}$ and $\hat{y'_p}$ to construct compound feature representations \new{$u$ and $u'$} both residing in the space $\mathbb{R}^{c \times h \times w}$. The aggregation process is a critical step that \new{blends} the information extracted from diverse intermediate feature maps. The aggregated compound feature representations $u$ and $u'$ encapsulate an amalgamation of \new{information} extracted from various layers, incorporating both local and global contextual cues. 
Before the aggregation occurs, a pre-processing step is applied to ensure consistency \new{in terms of $h$ and $w$} across different $\hat{y_p}$ and $\hat{y'_p}$.

\subsection{Loss Function}
The MLP heads ($g_\phi$ and $g_{\phi'}$) along with ($h_\phi$ and $h_{\phi'}$) serve as expansion components, projecting the global representations ($y$ and $y'$) and the aggregated compound representations ($u$ and $u'$) into higher-dimensional spaces \new{respectively}. This enables the subsequent hierarchical computation of loss functions. Specifically, $g_\phi$ and $g_{\phi'}$ transform $y$ and $y'$ into global embeddings denoted as $z$ $=$ $g_\phi(y)$ and $z'$ $=$ $g_{\phi'}(y')$, while $h_\phi$ and $h_{\phi'}$ map the compound representations into $m$ $=$ $h_{\phi}(u)$ and $m'$ $=$ $h_{\phi'}(u')$. \new{Motivated from~\cite{vicreg}, we defined the following two losses for global embeddings $z$ and $z'$ and multi-level embeddings $m$ and $m'$: }

\begin{equation}
\begin{split}
\mathcal{L_G} = \alpha(inv(Z,Z')) + \beta(var(Z)+var(Z')) \\ 
                + \gamma(cov(Z)+cov(Z'))
\label{ssl_lossa}
\end{split}
\end{equation}
\begin{equation}
\begin{split}
\mathcal{L_L} = \alpha(inv(M,M')) + \beta(var(M)+var(M')) \\
                + \gamma(cov(M)+cov(M'))
\label{ssl_lossb}
\end{split}
\end{equation}
\new{where,$Z$ and $M$ represent batches of global embeddings ($z$ and $m$), while following a similar analogy $Z'$ and $M'$ represent batches of corresponding multi-level embeddings ($z'$ and $m'$).}

The \new{overall} loss function for MLVICX is a composite of $\mathcal{L_G}$ and $\mathcal{L_L}$, as outlined in Eq.~\eqref{final_loss}.
\begin{equation}
\mathcal{L}_{mlca} = \mathcal{L_G} + \lambda \times \mathcal{L_L}
\label{final_loss}
\end{equation}
Here, $\lambda$ is the \new{balance factor}. 
The symbols $inv$,  $var$ and $cov$ are defined according to equation~\eqref{inv}, ~\eqref{var} and~\eqref{cov} respectively.

\begin{equation}
    inv(Z,Z') = \frac{1}{b} \sum_{i=1}^b \|z_i - z'_i \|^2_2 
    \label{inv}
\end{equation}
\new{The Eq.~\eqref{inv} calculates the euclidean distance between each pair of corresponding representations ($z_i$ and $z'_i$) in the batches Z and Z'.This quantifies how different the representations learned by the two branches are on average.
The $var(Z)$ term encourages the representations in batch $Z$ to have a variance for each dimension above a threshold $\Lambda$ which is set as 1. }
\begin{equation}
\begin{split}
var(Z) = \frac{1}{r} \sum_{j=1}^r max(0, \Lambda - (\sqrt{Var(z_{:j}) + \epsilon} )) \\
Var(z_{: j}) =\frac{1}{b-1} \sum_{i=1}^{b}(z_{ij}-\bar{z}_{j})^{2}, \quad \bar{z}_{j}=\frac{1}{b} \sum_{i=1}^{b} z_{ij}
\label{var} 
\end{split}
\end{equation}
Here, $b$ refers to batch size, and $r$ represents the dimensionality of the feature vector. The hyperparameters $\Lambda$ and $\epsilon$ are introduced to govern the target value of standard deviation and to ensure numerical stability, respectively. \new{The $cov(Z)$ term captures the covariance between different dimensions of a batch of representations $Z$ by taking the square of the off-diagonal components of the covariance matrix $C(Z)$. The off-diagonal elements of this matrix represent the covariances between different dimensions, indicating how these dimensions vary together across data points. Minimizing $cov(Z)$ encourages the representations to have low covariances between their dimensions, which can help to capture more independent and discriminative features.}

\begin{equation}
\begin{split}
C(Z) = \frac{1}{b-1} \sum_{i}^b (z_{i:} - \bar{z_{i:}} )(z_{i:} - \bar{z_{i:}})^T, \bar{z} = \frac{1}{b}\sum_{i}^b z_{i:}  \\cov(Z)=\frac{1}{r} \sum_{\ell \neq q} C(Z)_{\ell q}^{2}             
\label{cov}
\end{split}
\end{equation}
The ultimate aim of these regularization terms is to enhance the qualities of the embeddings: $inv(Z, Z')$ measures their similarity, $var(Z)$ enforces variability across the batch, and $cov(Z)$ emphasizes decorrelation between different dimensions of the embeddings. The coefficients $\alpha$, $\beta$, and $\gamma$ are scalar factors that weigh the contributions of the different loss terms. \new{Analogous to $Z$, the $inv$, $var$, and $cov$ terms applied to $M$ to enhance the quality of the multi-level embeddings.  $inv(M, M')$ measures the similarity between corresponding multi-level embeddings $m_i$ and $m'_i$ in batches $M $ and $M'$, respectively. Similarly, $var(M)$ enforces variability control across the batch, and $cov(M)$ encourages decorrelation between different dimensions of the multi-level embeddings.}

\section{Experimental Setup }
In this section, we detail our experiments, a description of the datasets, followed by the evaluation protocols, implementation details, and baselines. 
\subsection{Descriptions of Datasets}
\textbf{NIH Chest X-ray-14 Dataset:} In our pretraining phase, we \new{use} the NIH-Chest X-ray-14 dataset~\cite{wang2017chestx}. This dataset encompasses a vast collection of 112,120 X-ray images procured from 30,805 distinct subjects. Each image \new{comes} with one \new{or more} of 14 thorax-related diseases. We use the official 86,524 training samples, resized to dimensions of 224 x 224 during the pretraining phase, while 25,596 test samples are used for subsequent evaluations in downstream tasks.

\textbf{Vinbig-CXR Dataset:} The Vinbig-CXR dataset~\cite{nguyen2022vindr} serves as a substantial and diverse repository of chest X-ray images enriched with annotated labels. \new{With a collection of 18,000 view X-ray scans, along with detailed annotations for findings localization and thoracic disease classification}. Our study uses the Vinbig-CXR dataset to conduct classification tasks during downstream evaluations.

\textbf{SIIM-ACR Pneumothorax Dataset:}
The SIIM-ACR~\cite{siim} dataset serves as a benchmark for detecting pneumothorax, a condition characterized by air in the pleural cavity leading to lung collapse. Comprising a total of 12,047 chest radiographs, each paired with manually created segmentation masks specifically targeting pneumothorax. For our evaluations, we carefully curated an equal number of positive \new{(pneumothorax)} and negative \new{(healthy)} samples for segmentation tasks, allocating 20\% of the samples for validation \new{which is a standard practice}.

\textbf{RSNA Pneumonia:} The RSNA~\cite{rsna} dataset consists of 30k chest radiographs captured from a frontal view, each annotated as either healthy or pneumonia-positive. This dataset is employed to assess the adaptability of our pretrained representations for classification tasks across external datasets.

\subsection{Evaluation Protocols} To assess the \new{effectiveness} of the learned representations from our proposed approach, we conducted downstream tasks involving chest X-ray image classification and segmentation. These tasks were executed using only the backbone encoder obtained from pretraining.

Following established practices in the literature such as SimCLR~\cite{simclr}, BYOL~\cite{byol}, PCRLv2~\cite{pcrlv2}, and VICReg~\cite{vicreg},  we adopted two distinct evaluation protocols to showcase the robustness and efficiency of our framework: (1) fine-tuning and (2) frozen. For classification tasks, in both evaluation settings, we added a single linear layer classifier on top of the pretrained backbone CNN encoder while for the segmentation task, we adopted a U-Net architecture. In the fine-tuning evaluation protocol, we fine-tuned both the SSL pretrained backbone CNN encoder and the newly added linear classifier layer. Conversely, in the frozen evaluation protocol, the parameters of the backbone CNN encoder remain fixed, ensuring that only the linear layer's parameters are updated during downstream training. We presented results \new{for downstream training} on the test/validation dataset using various subsets (1\%, 10\%, 30\%, and All) of the training data.

For the segmentation task, we initialized the U-Net encoder's weights with the SSL pre-trained encoder's weights. During the fine-tuning setting, both the backbone encoder and the decoder are updated during downstream training, while under the frozen evaluation setting, only the decoder's weights are updated in downstream training. 
Notably, the U-Net decoder is initialized randomly. In both evaluation settings, we evaluated the representations learned by our proposed SSL method using different subsets (30\% and 100\%) of the \new{downstream} training data. The 1\% and 10\% subsets are omitted due to limited data availability. To effectively \new{measure} the performance of our method, we utilized the AUC score for classification tasks and the Dice score for segmentation tasks.

\subsection{Implementation Details}
MLVICX employs the ResNet18 architecture as the backbone encoder for the SSL pretraining phase. This training process encompasses a batch size of $64$ over a duration of 300 epochs. Within the MLVICX framework, the expander heads consist of three linear layers, each configured with a dimensionality of $2048$. These layers are augmented with ReLU activation functions and batch normalization. Our optimization strategy relies on the LARS (Layer-wise Adaptive Rate Scaling) algorithm, utilizing a learning rate of 3e-5 and applying a weight decay of $1e-5$. The hyperparameters $\alpha$, $\beta$, and $\gamma$ are directly adopted from VICReg~\cite{vicreg}. During downstream evaluations, we maintain a batch size of $64$ for classification tasks on the NIH dataset. For the Vinbig-CXR and RSNA datasets, a batch size of $16$ is used. For the classification tasks on the NIH dataset, we fine-tune with a learning rate of $1e-5$.
In contrast, for the Vinbig-CXR and RSNA datasets, the learning rate is adjusted to $1e-3$. For segmentation tasks across all datasets, we set the batch size to $32$, and the learning rate is fine-tuned to $3e-4$ for optimal performance. In all downstream tasks, we incorporate early stopping with patience of $10$ epochs, ensuring that the training process \new{stops} efficiently based on validation performance criteria. 

\subsection{Baselines}
Our comparative analysis thoroughly evaluated the proposed approach against various baseline methods that encompassed both supervised and SSL techniques across various categories. For supervised training from scratch (Rand. Init.), we trained the model using labeled data with our designated settings. In the SSL domain, we included SOTA contrastive and non-contrastive methods, SimCLR~\cite{simclr}, MoCov2~\cite{moco}, BYOL~\cite{byol}, SimSiam~\cite{simsiam}, Barlow-Twins (BT)~\cite{barlow}, and VICReg~\cite{vicreg}. We conducted the pretraining for these baselines ourselves, following their official implementations and keeping in alignment with our proposed approach's training protocol. Specifically, we conducted the pretraining for $300$ epochs using a batch size of $64$ without modifying their default hyperparameters or network configurations. \new{Further for comparison with} the specific domain of medical image analysis, we also incorporated SSL baselines tailored for medical images, namely PCRLv1~\cite{pcrlv1} and PCRLv2~\cite{pcrlv2}. We utilized the official publically \new{available implementations} of these methods. The pretraining is conducted on the official training dataset provided by the NIH \new{as mentioned in the previous section}.

\begin{table*}[h!]
\setlength{\tabcolsep}{4pt}
\centering
\caption{Evaluation of the proposed method alongside various baselines on the downstream classification task using the NIH dataset. The Table presents the test set results after three independent runs under the finetuning linear evaluation protocol on different subsets of training data (1\%, 10\%, 30\%, and 100\%), measured in terms of mean AUC scores. Class-wise AUC scores are reported for each medical condition, demonstrating MLVICX's consistent improvements across a diverse range of pathologies compared to other self-supervised and supervised baselines.}
\label{tab:nihclass}
\begin{tabular}{clccccccccccccccc}
\toprule
Data \% & Methods  & \rotatebox{90}{Atelectasis} & \rotatebox{90}{Cardiomegaly} & \rotatebox{90}{Consolidation} & \rotatebox{90}{Edema} & \rotatebox{90}{Effusion} & \rotatebox{90}{Emphysema} & \rotatebox{90}{Fibrosis} & \rotatebox{90}{Hernia} & \rotatebox{90}{Infiltration} & \rotatebox{90}{Mass} & \rotatebox{90}{Nodule} & \rotatebox{90}{Pleural\_Thick.} & \rotatebox{90}{Pneumonia} & \rotatebox{90}{Pneumothorax} & \rotatebox{90}{Mean} \\
\midrule
\multicolumn{1}{c}{\multirow{10}{*}{1\%}} & Rand. Init   &0.570  &0.627  &0.616  &0.677  &0.647  &0.478  &0.648  &0.533  &0.611  &0.517 &0.487  &0.528  &0.585  &0.513 &0.577 \\
\multicolumn{1}{c}{} & SimCLR  & 0.650 &0.647 &0.567 & 0.658 & 0.715 & 0.504 & 0.663 &0.646 &0.627 &0.592 &0.548 &0.623 &0.580  &0.632 & 0.621 \\
\multicolumn{1}{c}{} & MoCov2  &0.605  &0.618  &0.619  &0.651  &0.665  &0.535  &0.644  &0.556  &0.614  &0.572  &0.565  &0.583  &0.565  &0.605 & 0.602  \\
\multicolumn{1}{c}{} & BYOL  & 0.609 &0.631  &0.621  &0.657  &0.670  &0.543  &0.690 &0.583 &0.594 &0.567 &0.620 &0.609 &0.570 &0.589 & 0.610 \\
\multicolumn{1}{c}{} & SimSiam  &0.610  &0.611  &0.601  &0.641  &0.647  &0.524  &0.670  &0.603  &0.579  &0.572 &0.589  &0.619  &0.569 &0.608 & 0.603  \\
\multicolumn{1}{c}{} & Barlow-Twins  &0.619   &0.678  &0.620  &0.677  &0.672  &0.464  &0.664  &\textbf{0.668}  &0.624  &0.568  &0.501  &0.601  &0.578  &0.604 & 0.614  \\
\multicolumn{1}{c}{} & VicReg  &0.629  &0.652  &0.539  &0.689  &0.668  &0.467  &0.646  &0.609  &0.627  &0.583  &0.507  &0.576  &0.573  &0.562 & 0.600  \\
\multicolumn{1}{c}{} & Inter\_MoCo~\cite{intermoco}   &0.620  &0.580  &0.613  &0.608  &0.699  &0.532  &0.538  &0.521  &0.625  &0.570  &0.526  &0.590  &0.543  &0.630 & 0.601  \\
\multicolumn{1}{c}{} & PCRLv1  &0.627  &0.690  &0.589  &0.693  &0.704  &0.523  &\textbf{0.690}  &0.451  &0.628  &0.615  &0.552  &0.622  &0.582  &0.626 & 0.617   \\
\multicolumn{1}{c}{} & CAiD   &0.595  &0.565  &0.650 &0.614  &0.699  &0.480  &0.612  &0.476 &0.649  &0.573  &0.508  &0.592  &0.572  &0.646 & 0.594 \\
\multicolumn{1}{c}{} & PCRLv2&0.640  &\textbf{0.693 } &0.607  &\textbf{0.761}  &0.706  & \textbf{0.596} & 0.623 &0.438  &0.617  &\textbf{0.645}  &0.575  &0.622  & 0.595 & \textbf{0.673}  & 0.630  \\
\multicolumn{1}{c}{} & MLVICX  &\textbf{0.656}  &0.632  &\textbf{0.657}  &0.735  &\textbf{0.719}  &0.553  &\textbf{0.667}  &0.625&\textbf{0.638}  &0.556  &\textbf{0.596} &\textbf{0.638}  &\textbf{0.598}  & 0.635 & \textbf{0.642} \\
\midrule
\multirow{10}{*}{10\%} & Rand. Init   &0.650 &0.595  &0.671  &0.761  &0.692  &0.609  &0.703  &0.770  &0.624  &0.601  &0.586  &0.628  &0.604  &0.691 &0.656  \\
 & SimCLR   &0.675  &0.772  &0.690  &0.756  &0.714  &0.646  &0.707  &0.782  &0.640  &0.672  &0.575  &0.667  &0.629  &0.722 & 0.689  \\
 & MoCov2  & 0.665 &0.702  &0.640  &0.766  &0.719  &0.686  &0.657  &0.748  &0.661  &0.681  &0.601  &0.697  &0.646  &0.702 & 0.684  \\
 & BYOL  &0.635  &0.732  &0.643  &0.726  &0.701  &0.648  &0.679  &0.723  &0.672  &0.623  &0.654  &0.689  &0.669  &0.682 & 0.677 \\
 & SimSiam  &0.668  &0.679  &0.693  &0.765  &0.718  &0.651  &0.721  &0.773  &0.633  &0.593  &0.541  &0.627  &0.632  & 0.688 & 0.668\\
 & Barlow-Twins   &0.680  &0.765  &0.691  &0.776  &0.741  &0.675 & 0.689 &0.702  &0.711  &0.641  &0.656  &0.617  &0.658  &0.635  &0.718  \\
 & VicReg &0.671  &0.690  &0.685  &0.776  &0.722  &0.636  &0.726  &0.795  &0.660  &0.628  &0.621  &0.643  &0.631  &0.696  & 0.684 \\
 & Inter\_MoCo~\cite{intermoco}     &0.705  & \textbf{0.767} & 0.705 &0.804  &0.770  &0.680  &0.702  &0.597  &0.662  &0.704  &0.622  &0.674  &0.626  &0.727 & 0.696   \\
 & PCRLv1  &0.669  &0.702  &0.694  &0.785  &0.734  &0.635  &0.702  &0.751  &0.664  &0.646  &0.575  &0.657  &0.630  &0.705 & 0.682 \\
 \multicolumn{1}{c}{} & CAiD   & 0.701  &0.527  &0.693  &0.740  &0.760  &0.650  &0.647  &0.540  &0.665  &0.637  &0.605  &0.656  &0.588  &0.696 & 0.653 \\
 & PCRLv2  &\textbf{0.708}  &0.764  &\textbf{0.713}  &0.801  &0.785  &0.760  &0.735  &0.682  &0.676  &\textbf{0.699}  &0.670  &0.689  &0.654  &0.768 & 0.720 \\
 & MLVICX  &0.689  &0.757  &0.708  &\textbf{0.806}  &\textbf{0.788}  &\textbf{0.762}  &\textbf{0.762}  &\textbf{0.835}  &\textbf{0.677}  &0.695  &\textbf{0.675}  &\textbf{0.700}  &0.651  &\textbf{0.783} & \textbf{0.733} \\
\midrule
\multirow{10}{*}{30\%} & Rand. Init  &0.694  &0.791  &0.687  &0.778  &0.751  &0.707  &0.715  &0.740  &0.654  &0.674  &0.610  &0.691  &0.650  &0.749 &0.707  \\
 & SimCLR  &0.710  &0.825  &0.701  &0.793  &0.781  &0.723  &0.733  &0.728  &0.666  &0.685  &0.642  &0.709  &0.653  &0.773 & 0.722 \\
 & MoCov2  &0.704  &0.808  &0.693  &0.801  &0.705  &0.728  &0.749  &0.775  &0.673  &0.731  &0.649  &0.705  &0.655  & 0.774& 0.719 \\
 & BYOL  &0.698  &0.791  &0.691  &0.797  &0.759  &0.679  &0.743  &0.807  &0.666  &0.669  &0.630  &0.695  &0.642  &0.746& 0.716  \\
 & SimSiam &0.690  &0.783  &0.688  &0.793  &0.764  &0.672  &0.745  &0.832  &0.650  &0.672  &0.640  &0.685  &0.643  &0.746 & 0.713  \\
 & Barlow-Twins   &0.701  &0.835  &0.697  &0.791  &0.776  &0.695  &0.737  &0.758  &0.661  &0.703  &0.624  &0.700  &0.659  &0.750 & 0.719 \\
 & VicReg   &0.702  &0.807  &0.687  &0.809  &0.765  &0.712  &0.747  &0.779  &0.639  &0.691  &0.626  &0.701  &0.657  &0.762 & 0.718 \\
 & Inter\_MoCo~\cite{intermoco}    &0.701  &0.821  &0.708  &0.798  &0.773  &0.739  &0.734  &0.794  &0.651  &0.730  &0.658  &0.693  &0.637  &0.776 & 0.727  \\
 & PCRLv1   &0.689  &0.827  &0.681  &0.793  &0.768  &0.702  &0.744  &0.752  &0.670  &0.687  &0.640  &0.697  &0.645  &0.754 & 0.716 \\
 \multicolumn{1}{c}{} & CAiD   &0.721  &0.806  &0.685  &0.790  &0.786  &0.729  &0.732  &0.835  &0.656  &0.710  &0.666 &0.700  &0.647  &0.768 & 0.728 \\
 & PCRLv2 & 0.714  &0.822  &0.699  &0.804  &0.782  &0.722  &0.747  &0.809  &0.675  &0.714  &0.641  &0.714  &0.648  & 0.768 & 0.731  \\
 & MLVICX  &\textbf{0.730} &\textbf{0.859 } &\textbf{0.709}  &\textbf{0.816}  &\textbf{0.795}  &\textbf{0.821}  &\textbf{0.785}  &\textbf{0.843}  &\textbf{0.682}  &\textbf{0.754}  & \textbf{0.691} &\textbf{0.733}  &\textbf{0.692}  &\textbf{0.822} & \textbf{0.763}  \\
\midrule
\multirow{10}{*}{All} & Rand. Init   &0.721  &0.851  &0.709  &0.816  &0.792  &0.740  &0.751  &0.782  &0.684  &0.741  &0.657  &0.724  &0.652  & 0.779 &0.741\\
 & SimCLR  &0.731 &0.866  & 0.724 &0.828  &0.799  &0.781  &0.767  &0.869  &0.679  &0.730  &0.627  &0.735  &0.672  &0.808  &0.756 \\
 & MoCov2  &0.723  &0.839  &0.736  &0.809  &0.791  &0.749  &0.751  &0.851  &0.689  &0.734  &0.649  &0.731  &0.668  &0.798  &0.751 \\
 & BYOL   &0.728  &0.863  &0.721  &0.824  &0.796  &0.734  &0.760  &0.826  &0.672  &0.731  &0.669  & 0.727 &0.664  &0.788 &0.748\\
 & SimSiam  &0.728  &0.857  &0.717  &0.825  &0.797  &0.742  &0.752  &0.789  &0.683  &0.727  &0.640  &0.725  &0.677  &0.794 &0.744  \\
 & Barlow-Twins  &0.728  &0.861  &0.724  &0.825  &0.792  &0.771  &0.752  &0.841  &0.669  &0.736  &0.649  &0.717  &0.670  &0.785 &0.749  \\
 & VicReg  &0.730  &0.864  &0.721  & 0.823 &0.800  & 0.775 &0.760  &0.804  &0.678  &0.750  &0.667  &0.732  &0.684  &0.799&0.754   \\
 & Inter\_MoCo~\cite{intermoco}   &0.735  &0.864  &0.725  &0.832  &0.808  &0.814  &0.782  &0.848 &0.674  &0.766  &0.677  &0.740  &0.679  &0.811&0.765  \\
 & PCRLv1   &0.733  &0.861  &0.720  &0.822  &0.801  &0.795  &0.782  &0.848  &0.661  &0.745  &0.668  &0.728  &0.670  &0.806&0.758  \\
\multicolumn{1}{c}{} & CAiD   &0.751  &0.867  &0.730  &0.828  &0.818  &0.824 &0.776  &0.850  &0.696  &0.773 &0.707  &0.735  &0.708  &0.826 & 0.774 \\
 & PCRLv2  &0.730  &0.863  &0.720  &0.814  &0.802  &0.814  &0.777  &0.842  &0.684  &0.748  &0.682  &0.731  &0.687  &0.799  &0.761\\
 & MLVICX  &\textbf{0.759}  &\textbf{0.884}  &\textbf{0.736}  &\textbf{0.837}  &\textbf{0.821}  &\textbf{0.886}  &\textbf{0.809} &\textbf{0.870}  &\textbf{0.702}  &\textbf{0.788}  &\textbf{0.717}  &\textbf{0.758}  &\textbf{0.709}  &\textbf{0.854}  &\textbf{0.790}\\
\bottomrule
\end{tabular}
\end{table*}

\begin{table*}[htbp]
\setlength{\tabcolsep}{4pt}
\centering
\caption{\new{Performance evaluation of MLVICX and the considered baselines on NIH, Vinbig-CXR, and RSNA datasets, under both frozen and fine-tuning settings. Results are reported for different data percentages. The results for NIH and Vinbig-CXR datasets are reported in terms of AUC, while for the RSNA dataset, accuracy is used as the evaluation metric. These results represent the average performance across three independent runs, and SD is not shown due to minimal variability.}}
\label{tab:clasification}
\begin{tabular}{clcccccccccccc}
\toprule
Methods  & Data \% & \rotatebox{90}{Rand. Init} & \rotatebox{90}{SimCLR} & \rotatebox{90}{MoCov2} & \rotatebox{90}{BYOL} & \rotatebox{90}{SimSiam} & \rotatebox{90}{BT} & \rotatebox{90}{VicReg} & \rotatebox{90}{Inter\_MoCo~\cite{intermoco} } & \rotatebox{90}{PCRLv1} & \rotatebox{90}{CAiD} & \rotatebox{90}{PCRLv2} & \rotatebox{90}{MLVICX}  \\
\midrule
\multicolumn{1}{c}{\multirow{4}{*}{NIH (Frozen) }} & 1\% &0.516  &0.569  &0.561  &0.547  &0.539  &0.584  &0.587  &0.524 &0.566 &0.567 &0.579  &\textbf{0.601}   \\
\multicolumn{1}{c}{} & 10\% & 0.571 & 0.627 &0.621 &0.617 & 0.611 & 0.642 & 0.627 & 0.659 &0.620 &0.654 &0.643 &\textbf{0.661}   \\
\multicolumn{1}{c}{} & 30\% & 0.611 &0.676  &0.669  &0.663  &0.659  &0.660  &0.662  &0.689  &0.655 &0.683  &0.675  &\textbf{0.691}   \\
\multicolumn{1}{c}{} & ALL & 0.618 & 0.700 &0.689  &0.690  &0.688  &0.680  &0.673  &0.703 &0.672 &0.705 &0.691 &\textbf{0.710}  \\
\midrule
\multicolumn{1}{c}{\multirow{4}{*}{Vinbig-CXR (Frozen) }} & 1\% &0.568  &0.683  &0.643  &0.656  &0.650  &0.661  &0.653  &0.686  &0.695 &0.698 &0.598  &\textbf{0.709}   \\
\multicolumn{1}{c}{} & 10\% & 0.594 & 0.774 &0.736 &0.778 & 0.775 & 0.787 & 0.765 & \textbf{0.804} &0.772 &0.781 &0.719 &0.793   \\
\multicolumn{1}{c}{} & 30\% & 0.670 &0.826  &0.813  &0.821  &0.816  &0.827  &0.814  &0.829  &0.814 &0.824  &0.790   &\textbf{0.832}  \\
\multicolumn{1}{c}{} & ALL & 0.711 & 0.853 &0.855  &0.840  &0.841  &0.844  &0.832  &0.859 &0.850 &0.853 &0.851 &\textbf{0.861}  \\

\midrule
\multicolumn{1}{c}{\multirow{4}{*}{RSNA (Frozen) }} & 1\% &0.757&0.769 &0.768  &0.765  &0.762  &0.780  &0.767  &0.769  &0.768 &\textbf{0.790} &0.768  &0.783   \\
\multicolumn{1}{c}{} & 10\% &0.768 &0.805 &0.801  &0.784  &0.778  &0.808  &0.804 &\textbf{0.812} &0.805 &0.810 &0.803  &\textbf{0.812}  \\
\multicolumn{1}{c}{} & 30\% &0.770 &0.814 &0.808  &0.801  &0.798  &0.811  &0.809 &0.815  &0.802&0.814 &0.805  &\textbf{0.819}  \\
\multicolumn{1}{c}{} & ALL &0.782 &0.818 &0.812  &0.809  &0.805  &0.817  &0.812  & \textbf{0.820}&0.811 &\textbf{0.820} &0.811  &\textbf{0.820 }\\

\midrule
\multicolumn{1}{c}{\multirow{4}{*}{Vinbig-CXR (Finetune) }} & 1\% &0.722  &0.779  &0.771  &0.762  &0.758  &0.769  &0.754  &0.761  &0.776 &0.781 &0.759  &\textbf{0.808} \\
\multicolumn{1}{c}{} & 10\% & 0.847 & 0.877 &0.862 &0.858 & 0.859 & 0.875 & 0.865 & 0.894 &0.873&0.891 &\textbf{0.903 }&0.898   \\
\multicolumn{1}{c}{} & 30\% & 0.883 &0.899  &0.890  &0.889  &0.871  &0.903  &0.890  &0.919  &0.892 &0.919 &0.916  &\textbf{0.927}   \\
\multicolumn{1}{c}{} & ALL & 0.901 & 0.924 &0.920  &0.917  &0.915  &0.919  &0.919  &0.930 &0.914&0.938 &0.940 &\textbf{0.948}  \\
\midrule
\multicolumn{1}{c}{\multirow{4}{*}{RSNA (Finetune) }} & 1\% &0.770  &0.800  &0.794  &0.783  &0.781  &0.800  &0.795  &0.804  &0.789 &0.810 &0.801  &\textbf{0.815}   \\
\multicolumn{1}{c}{} & 10\% & 0.798 & 0.811 &0.813 &0.805 & 0.801 & 0.809 & 0.807 & 0.820 &0.799&0.818 &0.815 &\textbf{0.824}   \\
\multicolumn{1}{c}{} & 30\% & 0.800 &0.818  &0.820  &0.814  &0.816  &0.820  &0.817  &0.828  &0.813 &0.822 &0.823  &\textbf{0.832}   \\
\multicolumn{1}{c}{} & ALL & 0.810& 0.830  &0.827  &0.825  &0.821  &0.829  &0.827  &0.830 &0.820 &0.830&0.829 &\textbf{0.840}  \\

\bottomrule
\end{tabular}
\end{table*}

\section{Results and Discussion}
Table~\ref{tab:nihclass} presents the evaluation test set results of our proposed framework MLVICX, along with \new{considered} baselines, on the downstream classification task using the NIH dataset. \new{These results represent the average of three independent runs, where SD values are not displayed due to their small magnitudes}. This Table \new{highlights} the potential of SSL techniques to extract meaningful features from chest X-ray images, even when only a limited amount of labeled data is accessible for fine-tuning. The Table illustrates the performance of different methods when their pretrained representations are fine-tuned on subsets of the training data (1\%, 10\%, 30\%, and 100\%) using the linear evaluation protocol. The reported metrics include the mean AUC and class-wise AUC scores for different medical conditions. 

Observing the results, we can infer that MLVICX consistently outperforms the baselines across all data subsets. Particularly, for the 1\% data subset, the proposed approach achieves an AUC of 0.642, which is a notable improvement over the closest competing baseline with an AUC of 0.630. Notably, it achieves the highest mean AUC score among all methods, indicating its superior ability to extract meaningful representations. In the 10\% labeled data scenario, the proposed method consistently achieves the highest overall AUC scores with a mean AUC of 0.733. For the 30\% subset, MLVICX archived AUC 0f 0.763, which is 3\% higher than the closest baseline \new{that is PCRLv2. A similar trend is observed for the 100\% dataset. Note that PCRLv2 is a dedicated SSL approach for medical data.} 
Further, our method also exhibits remarkable class-wise improvements across various medical conditions. Notably, MLCA's exceptional performance highlights its suitability for advancing the field of medical image analysis.

Table~\ref{tab:clasification} comprehensively evaluates the MLVICX model's generalization capabilities on out-of-distribution medical imaging datasets, specifically the Vinbig-CXR and RSNA datasets. We evaluated the model's performance under frozen and fine-tuning settings within the linear evaluation protocol, with results presented on the test sets regarding the mean AUC score.\new{ We also assessed the performance of MLVICX on the NIH dataset, employing the frozen evaluation setting.} MLVICX consistently demonstrated impressive performance across all datasets in the frozen setting. Notably, even with a minimal 1\% of the data on the NIH dataset, MLVICX achieves an AUC score of 0.601, which \new{is better than all the considered baselines}. This highlights the model's ability to leverage its pre-trained knowledge effectively, making it a strong contender for scenarios with limited labeled data. MLVICX \new{consistently} outperformed other methods as the data percentage increased. 

For the Vinbig-CXR dataset, MLVICX maintains an advantage over its competitors in the frozen setting, with AUC scores consistently higher across different data percentages. Particularly, with only 1\% of the data, MLVICX achieved an AUC score of 0.709, showcasing its generalization ability. On the RSNA dataset, MLVICX maintains a \new{better performance} in the frozen setting \new{as well}. \new{Similarly for} the fine-tuning setting, MLVICX outperformed other methods across different data percentages on the Vinbig-CXR and RSNA datasets. Impressively, with 30\% of the data for fine-tuning on Vinbig-CXR, MLVICX achieved an AUC score of 0.927. In low-data scenarios, such as fine-tuning with only 1\% of the data, MLVICX maintained its effectiveness, achieving an AUC score of 0.815 on RSNA.

These findings highlight MLVICX's remarkable adaptability and consistency in medical image classification tasks. It effectively handles diverse data distributions, demonstrating \new{relatively better} AUC scores in both frozen and fine-tuning settings. This \new{shows} its potential suitability for practical clinical applications, particularly in scenarios characterized by limited and heterogeneous datasets.

\begin{table}[t!]
\setlength{\tabcolsep}{4pt}
\centering
\caption{Evaluation of semantic-segmentation SIIM-ACR Pneumothorax under both frozen and finetuning settings of SSL evaluation. We reported test set results for 30\% and 100\% data. The metric used is the dice score.}
\label{tab:segmentation}
          
\begin{tabular}{ccccc}
\toprule
\multirow{3}{*}{Methods} & \multicolumn{4}{c}{SIIM-ACR Pneumothorax (Dice)}  \\
&\multicolumn{2}{c}{Frozen} & \multicolumn{2}{c}{Finetune}  \\

& 30\% & All & 30\% &All\\   
\midrule
\multicolumn{1}{c}{Rand. Init}  &0.42 $\pm$ 0.00&0.48 $\pm$ 0.01&0.45 $\pm$ 0.04&0.51 $\pm$ 0.01  \\
\multicolumn{1}{c}{SimCLR}  &0.44 $\pm$ 0.03&0.50 $\pm$ 0.01&0.46 $\pm$ 0.00
&0.53 $\pm$ 0.03\\
\multicolumn{1}{c}{MoCov2} &0.43 $\pm$ 0.02&0.50 $\pm$ 0.02&0.46 $\pm$ 0.01&0.52 $\pm$ 0.02 \\
\multicolumn{1}{c}{BYOL}  &0.43 $\pm$ 0.02&0.49 $\pm$ 0.01&0.45 $\pm$ 0.04&0.52 $\pm$ 0.01\\
\multicolumn{1}{c}{SimSiam} &0.43 $\pm$ 0.02&0.49 $\pm$ 0.01&0.45 $\pm$ 0.04&0.52 $\pm$ 0.02\\
\multicolumn{1}{c}{Barlow-Twins} &0.44 $\pm$ 0.03&0.48 $\pm$ 0.03&0.46 $\pm$ 0.02&0.54 $\pm$ 0.03  \\
\multicolumn{1}{c}{VicReg} &0.43 $\pm$ 0.01&0.48 $\pm$ 0.02&0.46 $\pm$ 0.02&0.54 $\pm$ 0.04  \\
\multicolumn{1}{c}{Inter\_MoCo~\cite{intermoco} } &0.44 $\pm$ 0.02&0.53 $\pm$ 0.01&0.47 $\pm$ 0.02&0.55 $\pm$ 0.01 \\
\multicolumn{1}{c}{PCRLv1}&0.43 $\pm$ 0.02&0.48 $\pm$ 0.01&0.46 $\pm$ 0.02&0.52 $\pm$ 0.00  \\
\multicolumn{1}{c}{CAiD} &0.44 $\pm$ 0.02&0.55 $\pm$ 0.01&0.46 $\pm$ 0.01&0.56 $\pm$ 0.01  \\
\multicolumn{1}{c}{PCRLv2}&0.44 $\pm$ 0.01&0.56 $\pm$ 0.00&0.46 $\pm$ 0.04&0.57 $\pm$ 0.02\\
\multicolumn{1}{c}{MLVICX}&\textbf{0.45} $\pm$ \textbf{0.01}&\textbf{0.57} $\pm$ \textbf{0.00}&\textbf{0.48 $\pm$ 0.01}&\textbf{0.59} $\pm$ \textbf{0.01}\\
\bottomrule
\end{tabular}
\end{table}

Furthermore, we also measure the performance of the proposed SSL framework for segmentation in the downstream task. Table~\ref{tab:segmentation} presents the performance of MLVICX and considered baselines in the context of transfer learning for segmentation tasks, specifically focusing on pneumothorax segmentation within the SIIM-ACR dataset under both frozen and finetuning settings. The evaluation metric used here is the Dice coefficient, which measures the similarity between predicted and ground truth segmentations. Remarkably, MLVICX consistently outperforms all other methods on both the 30\% subset of the dataset and the entire dataset. Specifically for the frozen scenario in which models are evaluated directly after pertaining, MLVICX 
\new{shows comparatively better performance, demonstrating the significance of performance improvement achieved by MLVICX. }

In the fine-tuning setting, where pretrained models undergo additional training on the specific pneumothorax segmentation task, MLVICX's performance is consistently better. It achieves the highest dice coefficients for both the 30\% dataset subset and the entire dataset, with approximately 2\% of performance gain. 
\new{In these findings, it becomes evident that MLVICX excels not only in transfer learning but also maintains its superiority when fine-tuned for a specific downstream task.}

\subsection{Statistical Significance Analysis}

In this section, we \new{conducted} the statistical significance analysis to \new{understand} the performance of our proposed model, MLVICX, in comparison to the \new{best considered} baseline, PCRLv2. We run both models ten times \new{by randomly selecting the subsets specifically 1\%, 10\%, and 30\% of the samples from the NIH training set}, each involving fine-tuning of learned representations. Subsequently, both MLVICX and PCRLv2 \new{undergoes} fine-tuning of learned representations in each run. \textit{p}-values are calculated for each data percentage by using a paired t-test, considering the results from ten independent runs. With 1\% of the data, the \textit{p}-value is 6.99e-03, indicating a statistically significant difference in favor of MLVICX. For 10\% and 30\% data, the \textit{p}-value further diminished to 6.27e-04 and 2.42e-07 respectively.
The lower \textit{p}-value affirms the significant performance advantage of MLVICX over PCRLv2. \new{In all the experiments, the \textit{p}-value is considerably smaller than $0.01$, demonstrating consistent superiority of MLVICX.}

\subsection{Ablation Study}

\begin{table}[t!]
\centering
\caption{Results of an ablation study conducted on MLVICX across various data percentages on the NIH dataset. The study evaluates the impact of individual loss components, $\mathcal{L_G}$ and $\mathcal{L_L}$, in both fine-tuning and frozen settings. }
\begin{tabular}{cccccc}
\toprule
\multicolumn{1}{c}{\multirow{2}{*}{Data(\%)}} & \multicolumn{2}{c}{Loss} & \multicolumn{1}{c}{\multirow{2}{*}{1\%}} & \multicolumn{1}{c}{\multirow{2}{*}{10\%}} & \multicolumn{1}{c}{\multirow{2}{*}{30\%}} \\
\multicolumn{1}{c}{} & $\mathcal{L_G}$ & $\mathcal{L_L}$ & \multicolumn{1}{c}{} & \multicolumn{1}{c}{} \\
\midrule
\multirow{2}{*}{Finetune}  & \xmark & \cmark  & 0.637 & 0.727 & 0.759  \\
& \cmark  & \xmark &0.631  & 0.721 & 0.751   \\
& \cmark  & \cmark  & \textbf{0.642} & \textbf{0.733} & \textbf{0.763}\\
\midrule
\multirow{2}{*}{Frozen} & \xmark & \cmark  & 0.589 & 0.653  & 0.681 \\
& \cmark  & \xmark & 0.581 & 0.648 & 0.677   \\
& \cmark  & \cmark &\textbf{0.601}  &\textbf{0.661} & \textbf{0.691} \\

\bottomrule
\end{tabular}
\label{tab:ablation}
\end{table}

\begin{figure*}[t!]
\centering
\begin{tabular}{ccccccccccc}
\multicolumn{11}{l}{\includegraphics[width=1\linewidth]{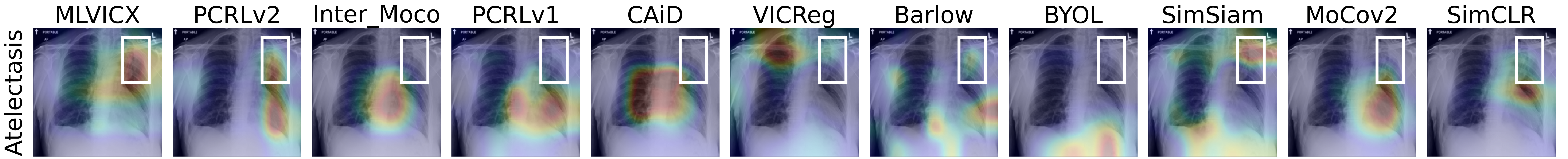}} \\
\includegraphics[width=1\linewidth]{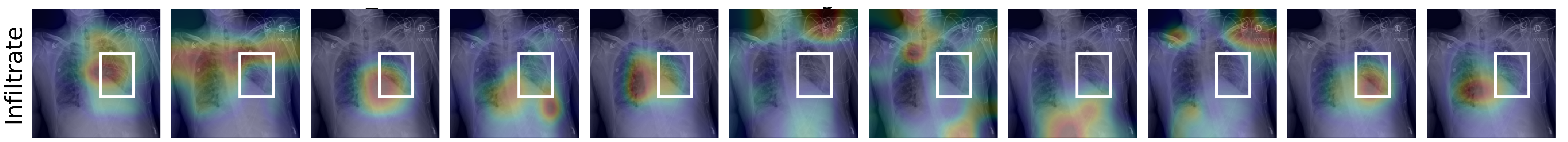} \\
\includegraphics[width=1\linewidth]{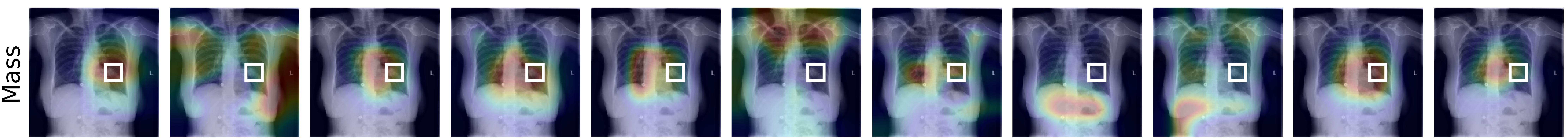} \\
\includegraphics[width=1\linewidth]{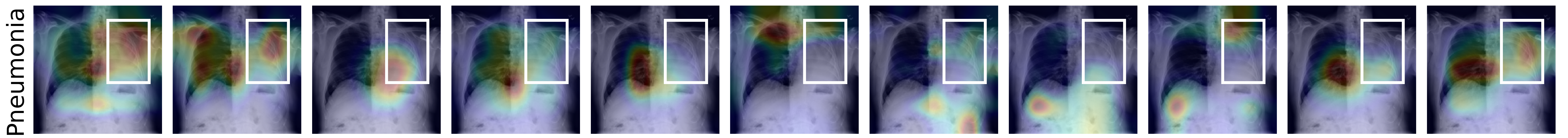} \\

\end{tabular}
\caption[short]{Diagnostic heatmaps generated by MLVICX and SSL baselines. The heatmaps represent model interpretations of chest X-ray images fine-tuned with 1\% of training samples from the NIH dataset. }
\label{fig_cam:gradcam}
\end{figure*}

In this section, we conducted an ablation study to examine the impact of the individual loss components within MLVICX. The model's performance is evaluated across various data percentages on the NIH dataset, both in the fine-tuning and frozen settings, by selectively disabling either the $\mathcal{L_G}$ loss or the $\mathcal{L_L}$ loss. Table~\ref{tab:ablation} summarizes the results of this study in terms of AUC, showcasing the significance of both $\mathcal{L_G}$ and $\mathcal{L_L}$ components in enhancing MLVICX's performance. When $\mathcal{L_L}$ is eliminated, indicating the exclusion of multi-level embeddings, the performance decline is slightly more prominent compared to the removal of $\mathcal{L_G}$. This suggests the importance of multi-level embeddings in the model's representation learning process.

\subsection{Qualitative Evaluation}
Fig.~\ref{fig_cam:gradcam} presents visual interpretations of chest pathology diagnoses for different conditions. We compare the performance of the proposed SSL framework MLVICX with all the considered SSL baselines. The heatmaps correspond to representations, fine-tuned with only 1\% of training samples from the NIH dataset. Notably, we observed that MLVICX consistently produces more interpretable diagnostic results when compared to the other methods. Specifically, in the case of pathologies like Mass, Atelectasis, and Infiltrate exhibit relatively better performance in comparison with the considered baselines. These pathologies often involve smaller lesion areas, making them more challenging to diagnose accurately. In the case of cardiomegaly, we find that most methods exhibit similar performance; however, MLVICX delivers more consistent and precise predictions. Even in the case of Pneumonia, MLVICX outperforms others. This analysis highlights MLVICX's effectiveness in handling a diverse range of pathologies.

\section{Conclusion}
This work proposes an SSL framework MLVICX that aggregates contextual representations from multiple intermediate layers. 
We demonstrate the performance of MLVICX on multiple chest X-ray datasets with multiple downstream tasks, including classification and segmentation. MLVICX can leverage limited labeled data while extracting meaningful features from chest X-rays through the combination of multilevel variance-covariance exploration. With its strong performance on various downstream tasks, MLVICX has the potential to become a valuable tool in various chest X-ray analysis applications. MLVICX offers a simple yet effective way to improve the generalization ability of the representations learned in transfer learning and when finetuned for specialized tasks.

\bibliographystyle{IEEEtran}
\bibliography{mlca}

\end{document}